\documentclass[twocolumn,english,showpacs,preprintnumbers,groupedaddress,amsmath,amssymb,aps,pra]{revtex4}
\usepackage[T1]{fontenc}
\usepackage[latin1]{inputenc}
\usepackage{float}
\usepackage{textcomp}
\usepackage{amsmath}
\usepackage{graphicx}
\usepackage{amssymb}
\usepackage{esint}
\usepackage{bm}

\makeatletter

\@ifundefined{textcolor}{}
{%
 \definecolor{BLACK}{gray}{0}
 \definecolor{WHITE}{gray}{1}
 \definecolor{RED}{rgb}{1,0,0}
 \definecolor{GREEN}{rgb}{0,1,0}
 \definecolor{BLUE}{rgb}{0,0,1}
 \definecolor{CYAN}{cmyk}{1,0,0,0}
 \definecolor{MAGENTA}{cmyk}{0,1,0,0}
 \definecolor{YELLOW}{cmyk}{0,0,1,0}
 }

\makeatother

\usepackage{babel}

\begin{document}

\title{From fast to slow light in a resonantly driven absorbing medium }

\author{Bruno Macke}

\author{Bernard  S\'{e}gard}

\email{bernard.segard@univ-lille1.fr}

\affiliation{Laboratoire de Physique des Lasers, Atomes et Mol\'{e}cules , CNRS et
Universit\'{e} Lille 1, 59655 Villeneuve d'Ascq, France}

\date{\today}
\begin{abstract}
We theoretically study the propagation through a resonant absorbing
medium of a time-dependent perturbation modulating the amplitude of
a continuous wave (cw). Modeling the medium as a two-level
system and linearizing the Maxwell-Bloch equations for the perturbation,
we establish an exact \emph{analytical} expression of the transfer
function relating the Fourier transforms of the incident and transmitted
perturbations. It directly gives the gain and the phase shift undergone
in the medium by a harmonic modulation. For the case of a pulse modulation,
it enables us to determine the transmission time of the pulse center-of-mass
(group delay), evidencing the relative contributions of the coherent
and incoherent (population) relaxation. We show that the group delay
has a negative value (fast light) fixed by the coherent effects when
the cw intensity is small compared to the saturation intensity and
becomes positive (slow light) when this intensity increases, before
attaining a maximum that cannot exceed the population relaxation time.
The analytical results are completed by numerical determinations of
the shape of the transmitted pulses in the different regimes. 
\end{abstract}


\maketitle

\section{Introduction}

Convincing demonstrations of fast light in a linear homogeneous medium
were performed in the 1980s by exploiting the steep anomalous dispersion
associated with a well-isolated, narrow and strong absorption line
of the medium, leading to negative values of the group velocity \cite{chu82,bs85}.
Ideally bell-shaped pulses can then propagate moderately distorted
in such a way that the maximum of the transmitted pulse occurs before
that of the incident pulse. Time-advances exceeding 0.5 times the
half-duration at half-maximum of the pulse-envelope have been so evidenced
in experiments involving a true shape detection of the latter \cite{bs85}.
See also \cite{tan03,bro08}. More generally, due to the causality
principle (originating the Kramers-Kronig relations), fast light is
expected every time that the carrier frequency of the pulses coincides
with the bottom of a well-marked \emph{dip} in the medium transmission
and the pulse distortion resulting from the first order variations
of the transmission and of the group velocity versus frequency cancels
when this coincidence is exact \cite{bm03}. The dip in the transmission
may be natural or created in various arrangements involving electromagnetically
induced absorption \cite{ak99,ak02}, stimulated Raman \cite{ste03}
or Brillouin \cite{gon05} scattering, etc. The literature on fast
light is abundant (for reviews, see, e.g., \cite{boy02,mil05,boy07,the08})
but it seems that the fractional pulse-advance with moderate distortion
reported in \cite{bs85} has not been overtaken. This is easily explained
by remarking that, since the transmission dip at the carrier frequency
must be well pronounced, there are spectral regions where the overall
transmission of the system is much larger. For obvious reasons of
noise (no matter its origin), instability and hypersensitivity to
small defects in the incident pulse-shape, the usable transmission-dynamics
cannot be too large and this limits the observable fast-light effects
in any linear system \cite{bm05}. The dynamics actually involved
in the fast-light experiments does not exceed 45 dB and is often much
lower. The situation is quite different in the slow-light experiments
where the medium transmission is \emph{maximal} at the carrier frequency
and can be very low on both sides of this frequency, with transmission-dynamics
exceeding 1000 dB \cite{kas95,cama07}.

In the present article, we study the propagation in an absorbing
medium of a continuous wave (cw) whose amplitude is pulse-modulated
with \emph{a low modulation index}. The medium is
modeled as a homogeneously-broadened two-level system (TLS) \cite{al87}
with a resonance frequency $\omega_{0}$ and a relaxation time $T_{1}$
($T_{2}$) for the population difference (the coherence). We assume
that the cw is on exact resonance and we are interested in the propagation
of the pulsed part of the wave (the pulse) whose optical spectrum
is centered on $\omega_{0}$ and maximum at this frequency. The pulse
may thus be seen as a continuous superposition of fields symmetrically
detuned from the cw field (sidebands) that probe the changes of the
TLS properties induced by the cw. The propagation in a strongly driven
TLS of two symmetrically detuned fields is a basic problem in nonlinear
optics \cite{sar78,sha10}. When the sidebands are associated with
an amplitude modulation, it has been experimentally evidenced \cite{sen63}
and theoretically demonstrated \cite{sar78,tvo91} that the transmission
of the modulation presents a maximum on resonance for strong enough
cw intensities. According to our general analysis, the pulse propagation
may be slow in such conditions. On the other hand, when the cw intensity
is small compared to the saturation intensity, the TLS behaves linearly
both for the cw and the pulse, the propagation of which is not modified
by the presence of the cw (fast light regime). Our main purpose is
to analyze how takes place the transition from a fast to a slow light
regime when the cw intensity increases. Our calculations take full
account of the depletion of the cw intensity during the propagation
and are made for arbitrary values of the relaxation times $T_{1}$
and $T_{2}$. The corresponding results (mainly analytical) are new
and it seems that the possibility of switching from fast to slow light
in a TLS by simply adding a resonant cw to the pulses has not been
previously considered.

The arrangement of our paper is as follows. In Sec. \ref{sec:Transfer-function-of},
we specify the system under consideration and we establish \emph{an
exact analytical expression} of its transfer function for the time-dependent
part of the field envelope. This expression directly gives
the gain and the phase shift undergone in the medium by a harmonic
modulation, examined in Sec.\ref{sec:HarmonicModulation}. The propagation
of pulses is studied in Sec. \ref{sec:Pulse-propagation}, with a
special attention paid to the transmission delay of their center-of-mass
(identified to the group delay). The analytical results are
completed by numerical determination of the envelope of the transmitted
pulses. We finally conclude in Sec. \ref{sec:Conclusion} by summarizing
the main results.

\section{Transfer function of the medium\label{sec:Transfer-function-of}}

As above mentioned we consider the case where the cw is exactly
resonant. The amplitude modulation is then an eigenmode of modulation
in the sense that that it is conserved during the propagation \cite{hil82,fra84,kra85}.
We denote $\ell$ the thickness of the medium, $\alpha$ its unsaturated
absorption coefficient on resonance for the intensity and $z$ ($0<z<\ell$)
the direction of propagation of the wave assumed to be plane and polarized
along the $x$-axis. We write the $x$-component of the electric field
as :
\begin{equation}
E_{x}(z,t)=\mathrm{Re}\left[\mathrm{e}^{i\omega_{0}t}\widetilde{E}(z,t)\right]\label{eq:1}
\end{equation}
 where $\widetilde{E}(z,t)$ is the slowly varying field-envelope.
As in all the following, $t$ is a \emph{local time}, that is the
real time minus $n_{0}z/c$, $c$ being the velocity of light in vacuum
and $n_{0}$ the refractive-index around $\omega_{0}$ of the eventual
host medium. Denoting $\mu$ the dipole matrix element for the transition
(chosen real), $R(z,t)=\frac{\mu\widetilde{E}(z,t)}{\hbar}$ the Rabi
frequency, $N(z,t)$ the population difference per volume unit ($N_{0}$
its value at equilibrium) and $\widetilde{P}(z,t)$ the envelope of
the electric polarization induced in the medium, it is convenient
to introduce the dimensionless quantities $D=\frac{N}{N_{0}}$, $P=i\frac{\widetilde{P}}{N_{0}\mu}\sqrt{\frac{T_{1}}{T_{2}}}$
and $E=\frac{\mu\widetilde{E}}{\hbar}\sqrt{T_{1}T_{2}}=R\sqrt{T_{1}T_{2}}$.
Note that all these quantities are real for the resonant case considered
here and that $I=E^{2}$ is the intensity normalized to the saturation
intensity. In such conditions, the Maxwell-Bloch (MB) equations governing
the system evolution take the simple form : 
\begin{equation}
\frac{\partial E}{\partial z}=-\frac{\alpha}{2}P\label{eq:2}
\end{equation}
\begin{equation}
T_{2}\frac{\partial P}{\partial t}=DE-P\label{eq:3}
\end{equation}
\begin{equation}
T_{1}\frac{\partial D}{\partial t}=-PE+(1-D)\label{eq:4}
\end{equation}
 These equations are easily solved when only the cw is present. In
the following we denote the corresponding values of $E,P,D$ and $I$
by the index $cw$. Combining Eqs. (\ref{eq:3}) and (\ref{eq:4}),
we find $D_{cw}(z)=1/\left[1+I_{cw}(z)\right]$ and $P_{cw}(z)=E_{cw}(z)/\left[1+I_{cw}(z)\right]$.
Replacing this result in Eq. (\ref{eq:2}), we get : 
\begin{equation}
\frac{\partial}{\partial z}\left[I_{cw}(z)+\ln I_{cw}(z)+\alpha z\right]=0\label{eq:5}
\end{equation}
\begin{equation}
I_{out}+\ln I_{out}=I_{in}+\ln I_{in}-\alpha\ell\label{eq:6}
\end{equation}
 where $I_{out}$ and $I_{in}$ are short hand notations of $I_{cw}(\ell)$
and $I_{cw}(0)$. When a small time-dependent perturbation $\delta E_{in}(t)$
(also real for an amplitude modulation) is added to the constant amplitude
$E_{in}=\sqrt{I_{in}}$ of the field of the incident cw, we search
for solutions of the MB equations under the form $E(z,t)=E_{cw}(z)+\delta E(z,t)$,
$D(z,t)=D_{cw}(z)+\delta D(z,t)$ and $P(z,t)=P_{cw}(z)+\delta P(z,t)$.
Taking into account the cw solution, we get at the first order of
perturbation 
\begin{equation}
\frac{\partial(\delta E)}{\partial z}=-\frac{\alpha}{2}\delta P\label{eq:7}
\end{equation}
\begin{equation}
T_{2}\frac{\partial(\delta P)}{\partial t}=E_{cw}\delta D+D_{cw}\delta E-\delta P\label{eq:8}
\end{equation}
\begin{equation}
T_{1}\frac{\partial(\delta D)}{\partial t}=-E_{cw}\delta P-P_{cw}\delta E-\delta D\label{eq:9}
\end{equation}
 These linear equations can be solved by introducing the Fourier transforms
$\Delta E(z,\Omega)$, $\Delta P(z,\Omega)$ and $\Delta D(z,\Omega)$
of $\delta E(z,t)$, $\delta P(z,t)$ and $\delta D(z,t)$ \cite{pap88}.
From Eqs (\ref{eq:8}) and (\ref{eq:9}), we get : 
\begin{equation}
\Delta P(z,\Omega)=\frac{\left[1+i\Omega T_{1}-I_{cw}(z)\right]\Delta E(z,\Omega)}{\left[1+I_{cw}(z)\right]\left[\left(1+i\Omega T_{1}\right)\left(1+i\Omega T_{2}\right)+I_{cw}(z)\right]}\label{eq:10}
\end{equation}
 Replacing this result in Eq.(\ref{eq:7}) and integrating on $z$,
we finally obtain the transfer function $H(\Omega)$ relating $\Delta E_{out}(\Omega)$
and $\Delta E_{in}(\Omega)$, the Fourier transforms of $\delta E_{out}(t)=\delta E(\ell,t)$
and $\delta E_{in}(t)$. It reads as : 
\begin{equation}
H(\Omega)=\exp\left(\intop_{0}^{\ell}f\left(z,\Omega\right)dz\right)\label{eq:11}
\end{equation}
 where 
\begin{equation}
f(z,\Omega)=\frac{\alpha}{2}\frac{I_{cw}(z)-\left(1+i\Omega T_{1}\right)}{\left[1+I_{cw}(z)\right]\left[\left(1+i\Omega T_{1}\right)\left(1+i\Omega T_{2}\right)+I_{cw}(z)\right]}\label{eq:12}
\end{equation}
 is the complex gain factor at the abscissa $z$. By means of Eq.(\ref{eq:5}),
the integration on $z$ in Eq.(\ref{eq:11}) can be transformed in
an integration on $I_{cw}$. We get : 
\begin{equation}
H(\Omega)=\exp\left[\frac{1}{2}\intop_{I_{in}}^{I_{out}}\frac{1+i\Omega T_{1}-I_{cw}}{I_{cw}\left[\left(1+i\Omega T_{1}\right)\left(1+i\Omega T_{2}\right)+I_{cw}\right]}dI_{cw}\right]\label{eq:13}
\end{equation}
 and finally 
\begin{multline}
H(\Omega)=\exp\left[\frac{1+i\frac{\Omega T_{2}}{2}}{1+i\Omega T_{2}}\ln\left(\frac{I_{in}+\left(1+i\Omega T_{1}\right)\left(1+i\Omega T_{2}\right)}{I_{out}+\left(1+i\Omega T_{1}\right)\left(1+i\Omega T_{2}\right)}\right)\right.\\ \left.-\frac{\ln\sqrt{I_{in}/I_{out}}}{\left(1+i\Omega T_{2}\right)}\right]\label{eq:14}
\end{multline}
 where $I_{out}$ and $I_{in}$ are related by Eq.(\ref{eq:6}). This
expression of $H(\Omega)$ is the central result of our paper. In
the limit $I_{in}\ll1$, $I_{out}\approx I_{in}\exp(-\alpha\ell)$
and $H(\Omega)$ is reduced to $\exp(-\frac{\alpha\ell}{2(1+i\Omega T_{2})})$
which, as expected, is the transfer function of a linear medium with
a Lorentzian absorption line \cite{bm03}. Quite generally, we see
that the poles of $H(\Omega)$ are in the half-plane $\mathrm{Im}\left(\Omega\right)>0$
and that $H\left(-\Omega\right)=H^{\star}\left(\Omega\right)$. This
ensures that the impulse response $h(t)$, inverse Fourier transform
of $H(\Omega)$, cancels for $t<0$ and is real, confirming that the
system is causal \cite{pap88} and that the amplitude modulation is
an eigenmode of modulation. At the first order of perturbation considered
here, we finally remark that \begin{equation}
I(z,t)=\left[E_{cw}(z)+\delta E(z,t)\right]^{2}\approx I_{cw}(z)+2\delta E(z,t)\sqrt{I_{cw}(z)}\label{eq:15}\end{equation}
 This means in particular that a quadratic detection, as currently
used in optics, will deliver a time dependent signal $\delta I_{out}(t)$
proportional to $\delta E_{out}(t)$ and that the instantaneous modulation
indices for the intensity and for the field are such that $\frac{\delta I(z,t)}{I_{cw}(z)}=2\frac{\delta E(z,t)}{E_{cw}(z)}$
in every point.

\section{Transmission of a harmonic modulation\label{sec:HarmonicModulation}}

\begin{figure}[h]
 \centering \includegraphics[width=75mm]{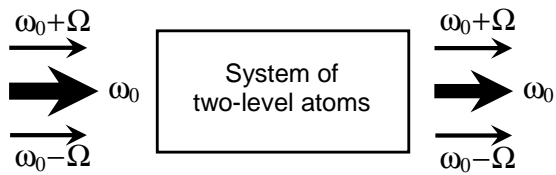} \caption{The driving wave (cw) at $\omega_{0}$ can modify the propagation
of the pair of sidebands at $\omega_{0}\pm\Omega$.\label{fig:Figure1}}

\end{figure}

\begin{figure}[h]
 \centering \includegraphics[width=80mm]{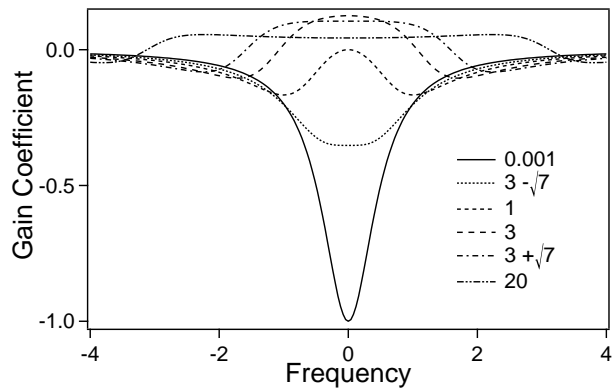} \caption{Gain coefficient $g(\Omega)$ in $\alpha/2$ units as a function of
$\Omega$ in $1/T_{1}$ units for $T_{2}=2T_{1}$. The different curves
are labeled by the corresponding value of the normalized intensity
$I_{cw}$ \label{fig:Gain-coefficient}.}

\end{figure}

\begin{figure}[h]
 \centering \includegraphics[width=80mm]{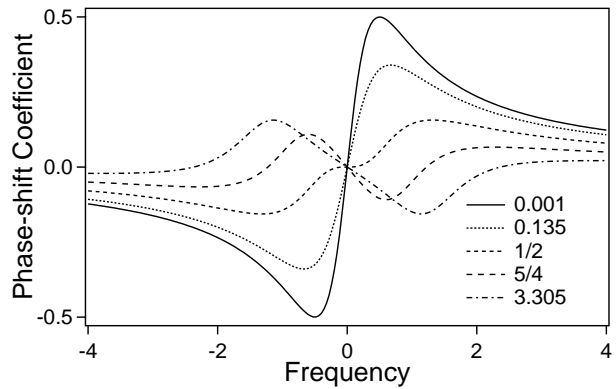} \caption{Same as Fig.\ref{fig:Gain-coefficient} for the phase-shift coefficient
$\varphi(\Omega)$ in $\alpha/2$ units.\label{fig:phase-shiftcoefficient}}

\end{figure}

When the amplitude-modulation is harmonic of frequency $\Omega$ ($\Omega>0$)
, the optical field has two sidebands at $\omega_{0}\pm\Omega$ that
both act as probes of the system driven at $\omega_{0}$ (Fig.\ref{fig:Figure1}).
The modulus $G(\Omega)$ and the argument $\Phi(\Omega)$ of $H(\Omega)$
are then respectively the gain and the phase-shift of the output amplitude-modulation
with respect to the input one. The gain coefficient $g(z,\Omega)=\mathrm{Re}\left[f(z,\Omega)\right]$
and the phase-shift coefficient $\varphi(z,\Omega)=\mathrm{Im}\left[f(z,\Omega)\right]$
are the corresponding quantities per unit length. Due to the depletion
of the cw intensity $I_{cw}$ during the propagation, these latter
quantities depend on $z$. Note that the expression of $g(z,\Omega)$
derived from Eq.(\ref{eq:12}) is consistent with the results given 
in \cite{sar78,tvo91}, also  as in \cite{sen63} for the particular
case $T_{1}=T_{2}$. Figures \ref{fig:Gain-coefficient} and \ref{fig:phase-shiftcoefficient}
show a set of profiles $g(\Omega)$ and $\varphi(\Omega)$ obtained
for $T_{2}=2T_{1}$ (purely radiative relaxation) with various cw
intensities. For $I_{cw}\ll1$ (fully linear case), we obviously retrieve
the gain and dispersion profiles $g(\Omega)=-\frac{\alpha}{2\left(1+\Omega^{2}T_{2}^{2}\right)}$
and $\varphi(\Omega)=\frac{\alpha\Omega T_{2}}{2\left(1+\Omega^{2}T_{2}^{2}\right)}$
, associated with a Lorentzian absorption line.

The gain coefficient (Fig.\ref{fig:Gain-coefficient}) keeps negative
as long as $I_{cw}<1$. It cancels at $\Omega=0$ when $I_{cw}=1$
(absolute cw intensity equal to the saturation intensity). Beyond,
it becomes positive in the frequency range $\Omega<\Omega_{g}=\sqrt{\left(I_{cw}^{2}-1\right)/\left(I_{cw}T_{1}T_{2}+T_{1}^{2}\right)}$,
attaining an absolute maximum $\alpha/16$ for $I_{cw}=3$. When $I_{cw}\gg1$,
$\Omega_{g}\approx R_{cw}$ , the Rabi frequency associated with the
cw field. We also note that, $g(\Omega)$ being an even function,
$g(0)=\frac{\alpha\left(I_{cw}-1\right)}{2\left(I_{cw}+1\right)^{2}}$
is always an extremum of the gain-coefficient. This extremum is flat
for $I_{cw}=b\pm\sqrt{b^{2}-T_{2}/T_{1}}$ with $b=3/2+T_{1}/T_{2}+T_{2}/2T_{1}$,
that is $I_{cw}=3\pm\sqrt{7}$ when $T_{2}=2T_{1}$. Despite some
common points, the profiles $g(\Omega)$ significantly differ from those obtained by using a probe 
independent of the driving field \cite{ha72,bo74,wu77,sar78,boy81}. They are a bit simpler, probably
because the amplitude modulation is an eigenmode of modulation (contrary
to the single sideband modulation). On the other hand, the refractive
index experienced by the optical probe field, well defined for the
single sideband case \cite{qua93}, is not defined for the
case of an amplitude modulation where the probe field has two frequency-components
at $\omega_{0}\pm\Omega$ that are not independent.

The phase-shift coefficient $\varphi(\Omega)$ (Fig. \ref{fig:phase-shiftcoefficient})
has a maximum $\varphi_{max}=\alpha/4$ (phase advance) at $\Omega=1/T_{2}$
for $I_{cw}\ll1$ and an absolute minimum $\varphi_{min}\approx-0.078\,\alpha$
(phase lag) at $\Omega\approx1.14/T_{1}$ for $I_{cw}\approx3.3$.
On the other hand, the slope $\frac{d\varphi}{d\Omega}\left|_{\Omega=0}\right.$
at the origin is also maximal for $I_{cw}\ll1$. This slope is half-maximum,
cancels and attains its (negative) minimum, for $I_{cw}$ respectively
equal to $0.135$, $1/2$ and $5/4$.

\begin{figure}[H]
 \centering \includegraphics[width=80mm]{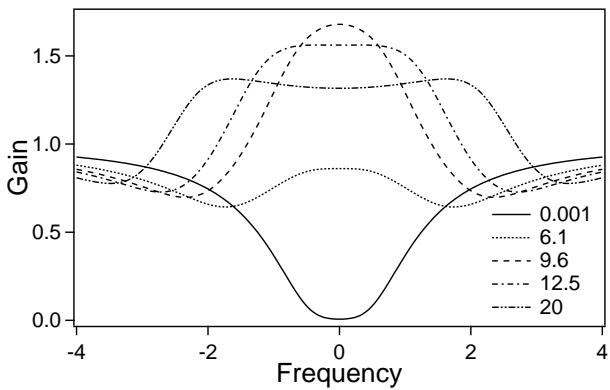}
\caption{Gain $G(\Omega)$ as a function of $\Omega$ in $1/T_{1}$ units for
$T_{2}=2T_{1}$ and $\alpha\ell=10$. The different curves are labeled
by the corresponding value of the normalized intensity $I_{in}$ .\label{fig:GainG}}

\end{figure}

\begin{figure}[h]
 \centering \includegraphics[width=80mm]{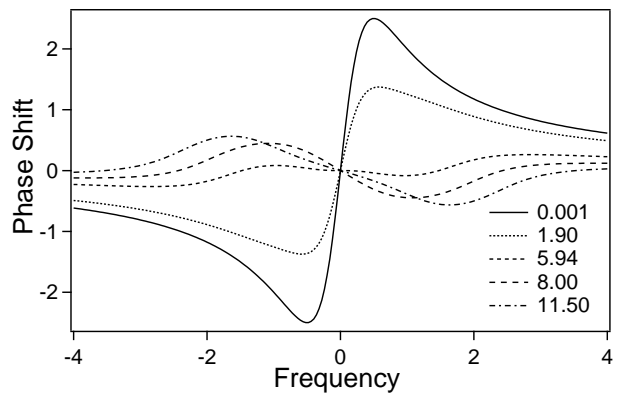} \caption{Same as Fig.\ref{fig:GainG} for the phase-shift $\Phi(\Omega)$.
The maximum and minimum phase-shift are respectively $\alpha\ell/4=2.5\,\mathrm{Rd}$
at $\Omega=1/T_{2}$ for $I_{in}\ll1$ and $-0.56\,\mathrm{Rd}$ at
$\Omega=1.68/T_{1}$ for $I_{in}\approx11.5$. \label{fig:PhaseShiftPHI.}}

\end{figure}

\begin{figure}[H]
 \centering \includegraphics[width=80mm]{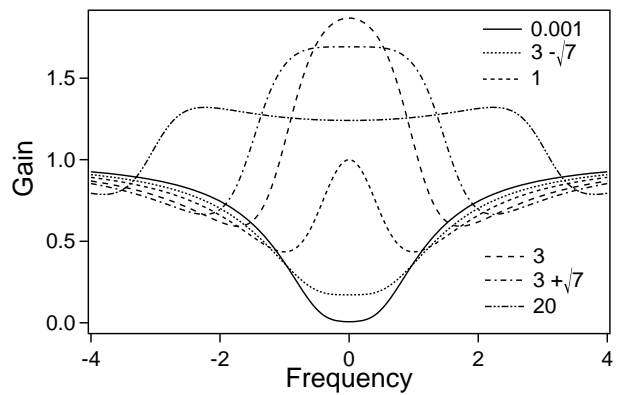} \caption{Same as Fig.\ref{fig:GainG} when the depletion of the driving wave
is not taken into account.\label{fig:GainSansDepletion}}

\end{figure}

Representative profiles of the (overall) gain $G(\Omega)$ and phase-shift
$\Phi(\Omega)$ are shown Fig.\ref{fig:GainG} and Fig.\ref{fig:PhaseShiftPHI.}
for $\alpha\ell=10$, with various intensities $I_{in}$ chosen according
to criteria analogue to those used Fig.\ref{fig:Gain-coefficient}
and Fig.\ref{fig:phase-shiftcoefficient}. The optical thickness retained
is of the order of that actually used in \cite{bs85}. When $I_{in}\ll1$,
$G(\Omega)=\exp\left[-\frac{\alpha\ell}{2\left(1+\Omega^{2}T_{2}^{2}\right)}\right]$
and $\Phi(\Omega)=\frac{\alpha\ell\Omega T_{2}}{2\left(1+\Omega^{2}T_{2}^{2}\right)}$.
Except for this fully linear case, Eqs. (\ref{eq:11}) and (\ref{eq:12})
show that $G(\Omega)$ and $\Phi(\Omega)$ are strongly affected by
the depletion of the cw intensity in the medium, often not taken into
account in the literature. This point is illustrated Fig.\ref{fig:GainSansDepletion}
where we give the profile $G(\Omega)=\exp\left[\ell g(\Omega)\right]$
which would be obtained by neglecting the cw depletion, i.e. by taking
$I_{cw}(z)=I_{in}$ everywhere. The depletion obviously forces one
to use larger incident cw intensities to obtain similar gain profiles
and the maximum gain is smaller. The difference is more important
when the optical thickness $\alpha\ell$ is very large. From Eqs.
(\ref{eq:6}) and (\ref{eq:14}), we find that the maximum gain then
tends to $\sqrt{\alpha\ell}/2$ whereas it equals $\exp\left[\alpha\ell/16\right]$
when the depletion is neglected. There are obviously similar effects
on the phase-shift $\Phi(\Omega)$. %

The previous results take a very simple form in the limit case where
$T_{2}$ is negligible compared to $T_{1}$. The MB equations are
then reduced to the rate equations \cite{al87} that provide a good
description of the propagation of light in numbers of saturable absorbers
\cite{sel67,sel70,zap06,pir07,bm08,sel09}. Putting $T_{2}\approx0$
in Eqs. (\ref{eq:12}) and (\ref{eq:14}), we get 
\begin{equation}
f(z,\Omega)=-\frac{\alpha}{2\left(1+I_{cw}\right)}+\frac{\alpha I_{cw}}{\left(1+I_{cw}\right)^{2}}\left(\frac{1}{1+i\frac{\Omega T_{1}}{1+I_{cw}}}\right)\label{eq:17}
\end{equation}
\begin{equation}
H(\Omega)=\sqrt{\frac{I_{out}}{I_{in}}}\left(\frac{1+I_{in}+i\Omega T_{1}}{1+I_{out}+i\Omega T_{1}}\right)\label{eq:18}
\end{equation}
 Apart from the term $-\frac{\alpha}{2\left(1+I_{cw}\right)}$ independent
of $\Omega$, the complex gain factor $f(z,\Omega)$ is identical
to that of a Lorentzian gain line of half-width at half maximum $\left(1+I_{cw}\right)/T_{1}$
with a gain-coefficient on resonance $\alpha I_{cw}/\left(1+I_{cw}\right)^{2}$
for the amplitude. During the propagation in the medium, the \emph{modulation}
\emph{index} is magnified whereas the phase of the modulation is delayed
\cite{sel71,hil83,bm08,sel09}. The magnification $K(\Omega)$ of
the modulation index and the phase lag $L(\Omega)=-\Phi(\Omega)$
for both the amplitude and the intensity modulations are easily deduced
from Eq.(\ref{eq:18}). We get : 
\begin{equation}
K(\Omega)=G(\Omega)\sqrt{\frac{I_{in}}{I_{out}}}=\sqrt{\frac{\left(1+I_{in}\right)^{2}+\Omega^{2}T_{1}^{2}}{\left(1+I_{out}\right)^{2}+\Omega^{2}T_{1}^{2}}}\label{eq:19}
\end{equation}
\begin{equation}
L(\Omega)=\tan^{-1}\left(\frac{\Omega T_{1}\left(I_{in}-I_{out}\right)}{\left(1+I_{in}\right)\left(1+I_{out}\right)+\Omega^{2}T_{1}^{2}}\right)\label{eq:20}
\end{equation}
 The phase lag, always positive for this {}``incoherent\textquotedblright{}
case, attains its maximum $L_{max}=\tan^{-1}\left[\left(I_{in}-I_{out}\right)/2\sqrt{\left(1+I_{in}\right)\left(1+I_{out}\right)}\right]$
for $\Omega T_{1}=\sqrt{\left(1+I_{in}\right)\left(1+I_{out}\right)}$.
We remark that $L_{max}<\pi/2$, the upper limit being approached
when $I_{in}\gg1$ and $I_{out}\ll1$, that is for $\alpha\ell\rightarrow\infty$.
Consequently the time delay of the output modulation can never exceed
a quarter of the modulation period $T=2\pi/\Omega$ (about an eighth
for $\alpha\ell=10$). Since the work reported in \cite{big03}, numerous
slow-light experiments performed in saturable media have been analyzed
by invoking hole burning via coherent population oscillations (CPO),
resulting in a reduction of the group velocity. In fact, a more direct
analysis is provided by the basic model of saturable absorber \cite{zap06}
and, as shown in detail in \cite{sel09}, the signals observed in
most CPO experiments can be perfectly reproduced by means of Eqs.
(\ref{eq:19}) and (\ref{eq:20}), eventually extended to take into
account inhomogeneous effects.

\section{Pulse propagation\label{sec:Pulse-propagation}}

\begin{figure}[h]
 \centering \includegraphics[width=80mm]{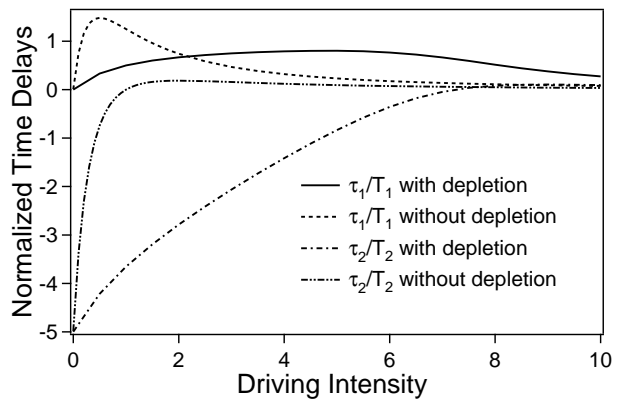} \caption{Normalized coherent and incoherent contributions to the group delay
as a function of the incident driving intensity $I_{in}$, depending
whether the cw depletion ($\alpha\ell=10$) is taken into account
or not.\label{fig:NormalizedDelays}}

\end{figure}

Strictly speaking a harmonic modulation does not contain any information
and, e.g., a time-delay of $T/8$ as considered at the end of the
previous section can also be seen as a time-advance of $7T/8$. Unambiguous
demonstrations of fast or slow light require to use pulses of finite
duration and energy. An important parameter to characterize the propagation
is then the transmission-time of the center-of-mass of the pulse envelope.
Following the use in signal theory \cite{pap88}, we define the center-of-mass
of a signal $y(t)$ of normalized area as $\intop_{-\infty}^{\infty}ty(t)dt$.
A direct application of the moment theorem \cite{pap88} shows then
that the transmission-time of the center-of-mass of the pulse envelope
is equal to the group delay $\tau_{g}=-\frac{d\Phi}{d\Omega}\left|_{\Omega=0}\right.$,
\emph{whatever the pulse distortion may be }\cite{bm03}. On the other
hand $-\frac{d\varphi}{d\Omega}\left|_{\Omega=0}\right.$ appears
as the transmission delay or group delay per unit length, equal in
our local time picture to $1/v_{g}-n_{0}/c\approx1/v_{g}$, where
$v_{g}$ is the group velocity. We emphasize that, due to the cw depletion,
$v_{g}$ is not uniform. From Eq.(\ref{eq:14}), we get $\tau_{g}=\tau_{1}+\tau_{2}$ with 
\begin{equation}
\tau_{1}=T_{1}\left(\frac{1}{I_{out}+1}-\frac{1}{I_{in}+1}\right)\label{eq:21}
\end{equation}
\begin{equation}
\tau_{2}=T_{2}\left(\frac{1}{I_{out}+1}-\frac{1}{I_{in}+1}-\ln\sqrt{\frac{1+1/I_{out}}{1+1/I_{in}}}\right)\label{eq:22}
\end{equation}
 $\tau_{1}$ ($\tau_{2}$), proportional to $T_{1}$ ($T_{2}$), may
be considered as the contribution to the group delay of the incoherent
(coherent) effects. We see that the incoherent or population effects
always lead to a delay whereas the coherent ones mainly lead to a
much larger time-advance (Fig.\ref{fig:NormalizedDelays}), at least
when $T_{2}$ is comparable to $T_{1}$ (the coherent effects obviously
disappear in a saturable absorber where $T_{2}\approx0$). %

Before commenting further on Eqs. (\ref{eq:21}) and (\ref{eq:22}),
consider the simpler results obtained when the depletion of the driving
wave is neglected ($I_{cw}=I_{in}$ everywhere). From Eq.(\ref{eq:12}),
we then get : 
\begin{equation}
\tau_{1}=T_{1}\frac{\alpha\ell I_{in}}{\left(1+I_{in}\right)^{3}}\label{eq:23}
\end{equation}
\begin{equation}
\tau_{2}=-T_{2}\frac{\alpha\ell\left(1-I_{in}\right)}{2\left(1+I_{in}\right)^{3}}\label{eq:24}
\end{equation}
 $\tau_{1}/T_{1}$, negligible for $I_{in}\ll1$, is maximum for $I_{in}=1/2$
whereas $\tau_{2}/T_{2}$ starts from the large negative value $-\alpha\ell/2$
before attaining a small positive maximum for $I_{in}=2$. The total
group delay cancels for $I_{in}=T_{2}/\left(T_{2}+2T_{1}\right)$
and is maximum for $I_{in}=\left(2T_{2}+T_{1}\right)/\left(T_{2}+2T_{1}\right)$
that is respectively for $I_{in}=1/2$ and $I_{in}=5/4$ when the
relaxation is purely radiative (see Fig.\ref{fig:phase-shiftcoefficient}).
As expected, taking into account the depletion of the cw intensity
does not modify the delays for $I_{in}\ll1$ but shifts the curves
$\tau_{1}(I_{in})$ and $\tau_{2}(I_{in})$ to larger $I_{in}$, dramatically
spreads them and reduces the amplitude of their maximum (Fig.\ref{fig:NormalizedDelays}).
Equation (\ref{eq:21}) shows that $\tau_{1}$ cannot overtake $T_{1}$,
this value being approached when $I_{in}\gg1$ and $I_{out}\ll1$,
that is for $\alpha\ell$ extremely large \cite{bm08}. An estimate
of the maximum of $\tau_{1}$ for large but realistic values of $\alpha\ell$
can be obtain by anticipating that it is attained for $I_{out}\ll1$.
From Eq.(\ref{eq:6}) we get then $I_{out}\approx I_{in}\exp\left(I_{in}-\alpha\ell\right)$
and 
\begin{equation}
\tau_{1}\approx T_{1}\left(\frac{1}{I_{in}\exp\left(I_{in}-\alpha\ell\right)+1}-\frac{1}{I_{in}+1}\right)\label{eq:25}
\end{equation}
 The maximum is obtained for $\left(I_{in}+1\right)^{3}\approx\exp\left(\alpha\ell-I_{in}\right)$.
For $\alpha\ell=10$ (Fig.\ref{fig:NormalizedDelays}), we find $I_{in}\approx4.8$
and, putting this value in Eq.(\ref{eq:25}), $\max\left(\tau_{1}/T_{1}\right)\approx0.80$
in perfect agreement with the exact value. By the same method we find
$\max\left(\tau_{1}/T_{1}\right)\approx0.995$ for $\alpha\ell=200$.
An upper bound to the total group delay can be obtained by using the
method of the Lagrange multipliers, perfectly adapted to the search
of a maximum under constraint, here fixed by Eq.(\ref{eq:6}). Following
the procedure exactly as exposed in the original work of Lagrange
\cite{la88}, we find an extra relation between $I_{in}$ and $I_{out}$
at the maximum of $\tau_{g}$, namely 
\begin{equation}
\frac{I_{in}}{I_{out}}=\left(\frac{I_{in}+1}{I_{out}+1}\right)^{3}\left(\frac{2T_{1}+T_{2}\left(1-1/I_{out}\right)}{2T_{1}+T_{2}\left(1-1/I_{in}\right)}\right)\label{eq:26}
\end{equation}
 A numerical exploration shows that the upper bound to the group delay
is attained when $I_{in}$ is extremely large whereas $I_{out}$ keeps
finite. Equation (\ref{eq:26}) takes then the form 
\begin{equation}
\left(1+I_{out}\right)^{3}\approx I_{in}^{2}\left[I_{out}-T_{2}/\left(2T_{1}+T_{2}\right)\right]\label{eq:27}
\end{equation}
 which can be verified if and only if $I_{out}\approx T_{2}/\left(2T_{1}+T_{2}\right)$.
Injecting this result in the general expression of $\tau_{g}$, we
finally find : 
\begin{equation}
\sup\left[\max\left(\tau_{g}\right)\right]=T_{1}+\frac{T_{2}}{2}-T_{2}\ln\sqrt{2\left(1+\frac{T_{1}}{T_{2}}\right)}\label{eq:28}
\end{equation}
 When $T_{2}\approx0$ (saturable absorber), we obviously retrieve
the upper bound $T_{1}$ for $\tau_{g}$ whereas this upper bound
is $\left(2-\ln3\right)T_{1}\approx0.90T_{1}$ when the relaxation
is purely radiative ($T_{2}=2T_{1}$). For the latter case $\max\left(\tau_{g}/T_{1}\right)$
is only $0.68$ for $\alpha\ell=10$, significantly below its upper
limit, and raises to $0.89$ for $\alpha\ell=200$.

As before mentioned, the group delay $\tau_{g}$ is the transmission-delay
of the pulse center-of-mass, as large as the pulse distortion may
be. On the other hand, the envelope $\delta E_{out}(t)$ of
the transmitted pulse is simply the inverse Fourier transform of $H(\Omega)\Delta E_{in}(\Omega)$.
When the duration $\tau_{p}$ of the incident pulse is long enough,
$\Delta E_{in}(\Omega)$ is concentrated around $\Omega=0$ where
$H(\Omega)\approx\sqrt{I_{in}/I_{out}}exp(-i\Omega\tau_{g})$. We
then get $\delta E_{out}(t)=\sqrt{I_{in}/I_{out}}\delta E_{in}(t-\tau_{g})$.
The whole envelope is multiplied by $\text{\ensuremath{\sqrt{I_{in}/I_{out}}}}$
and time shifted by $\tau_{g}$, without any distortion. Strictly
speaking, this solution is only valid for a very long pulse and is
not really interesting insofar as the time delay is then negligible
compared to the pulse duration. For finite $\tau_{p}$, there is always
some pulse-distortion whose importance depends in particular on the
transmission dynamics (as defined in the introduction). If, as usual,
the incident pulse is bell-shaped and symmetric with a maximum at
$t=0$, the envelope of the transmitted pulse will be dissymmetric
with a maximum at a time $\tau_{m}$ such that $\tau_{m}/\tau_{g}<1$,
its center-of-mass keeping exactly at the time $\tau_{g}$ no matter
the pulse duration. The challenge in the slow or fast light experiments
is to obtain a fractional delay or advance $\left|\tau_{m}\right|/\tau_{p}$
as large as possible, with moderate distortion. $\tau_{m}$ and $\tau_{g}$
are then comparable. The following figures are obtained by using
standard techniques of Fast Fourier Transform. The incident pulse
is Gaussian and $\tau_{p}$ is its half-duration at half-maximum.

\begin{figure}[h]
 \centering \includegraphics[width=80mm]{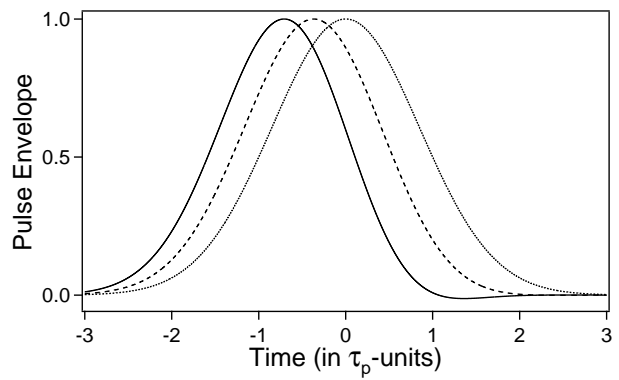} \caption{Normalized pulse envelopes obtained for $\alpha\ell=10$ and $T_{2}=2T_{1}$,
when the group \emph{advance} $-\tau_{g}$ is maximum (solid line)
and half-maximum (dashed line). $\tau_{p}$ is chosen such that the
fractional advance of the pulse maximum $-\tau_{m}/\tau_{p}$ is significant
whereas the distortion remains moderate ($\tau_{p}/T_{1}=12.5$).
For the first case, $\tau_{m}/\tau_{g}=0.88$ and $-\tau_{m}/\tau_{p}=0.71$.
For the second case $\tau_{m}/\tau_{g}=0.91$ and $-\tau_{m}/\tau_{p}=0.37$.
The envelope of the incident pulse is given for reference (dotted
line).\label{fig:Avance}}

\end{figure}

\begin{figure}[h]
 \centering \includegraphics[width=80mm]{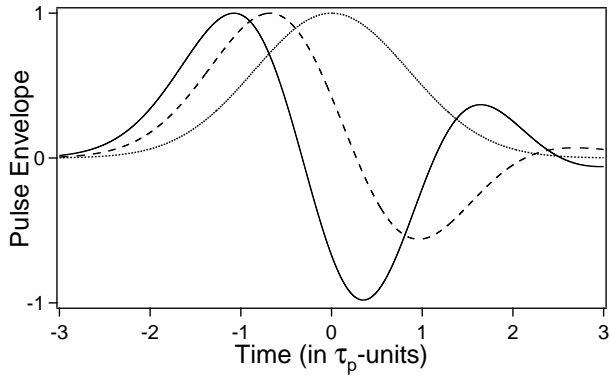} \caption{Same as Fig.8 when the pulse durations are those maximizing the fractional
advance $-\tau_{m}/\tau_{p}$ . For the solid line $\tau_{p}/T_{1}=4.1$
, $\tau_{m}/\tau_{g}=0.44$ and $-\tau_{m}/\tau_{p}=1.1$. For the
dashed line $\tau_{p}/T_{1}=2.8$ , $\tau_{m}/\tau_{g}=0.38$ and
$-\tau_{m}/\tau_{p}=0.68$.\label{fig:AvanceDistortion}}

\end{figure}

\begin{figure}[h]
 \centering \includegraphics[width=80mm]{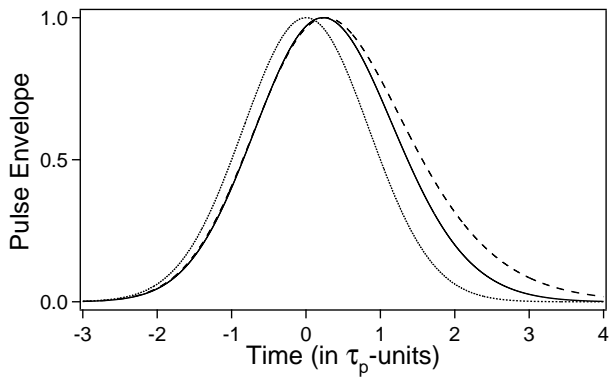} \caption{Normalized pulse envelopes obtained for $\alpha\ell=10$ and $T_{2}=2T_{1}$,
when the group \emph{delay} $\tau_{g}$ is maximum. For the solid
line $\tau_{p}/T_{1}=2.27$ , $\tau_{m}/\tau_{g}=0.77$ and $\tau_{m}/\tau_{p}=0.23$.
For the dashed line $\tau_{p}/T_{1}=1.44$ , $\tau_{m}/\tau_{g}=0.56$
and $\tau_{m}/\tau_{p}=0.27$. The envelope of the incident pulse
is given for reference (dotted line).\label{fig:RetardT2=00003D2T1}}

\end{figure}

Figures \ref{fig:Avance}, \ref{fig:AvanceDistortion} and \ref{fig:RetardT2=00003D2T1}
show typical shapes of the transmitted pulse for $T_{2}=2T_{1}$ (radiative
relaxation) and $\alpha\ell=10$. %

Figures \ref{fig:Avance} and \ref{fig:AvanceDistortion}
are obtained for $I_{in}=0.001$ and $I_{in}=1.9$ leading respectively
to a group \emph{advance} $-\tau_{g}$ nearly equals
to its maximum $\alpha\ell T_{2}/2$ and the half of this value. The
corresponding transmission dynamics is respectively $43\,\mathrm{dB}$
and $26\,\mathrm{dB}$. On Fig.\ref{fig:Avance}, $\tau_{p}$ has been
chosen in order that the fractional advance $-\tau_{m}/\tau_{p}$
is significant (respectively $0.71$ and $0.37$) and the distortion
keeps moderate ($\tau_{m}/\tau_{g}$ respectively equals $0.88$ and
$0.91$). When the pulse duration is shortened, the fractional advance
increases but not as much as one could expect (respectively
up to $1.1$ and $0.68$) and this is paid by a dramatic distortion
of the transmitted pulse (Fig.\ref{fig:AvanceDistortion}). %

Figure \ref{fig:RetardT2=00003D2T1} is obtained for $I_{in}\approx8.0$,
leading to the largest normalized group \emph{delay} ($\tau_{g}/T_{1}\approx0.68$).
The corresponding transmission dynamics is small ($7\,\mathrm{dB}$).
This explains in part that, even when the pulse duration is optimized
(dashed line), the fractional delay $\tau_{m}/\tau_{p}$ does not
exceed $0.27$. Another reason is that the coherent relaxation negatively
contributes to the group delay (see Fig.\ref{fig:NormalizedDelays}).
\begin{figure}[H]
 \centering \includegraphics[width=80mm]{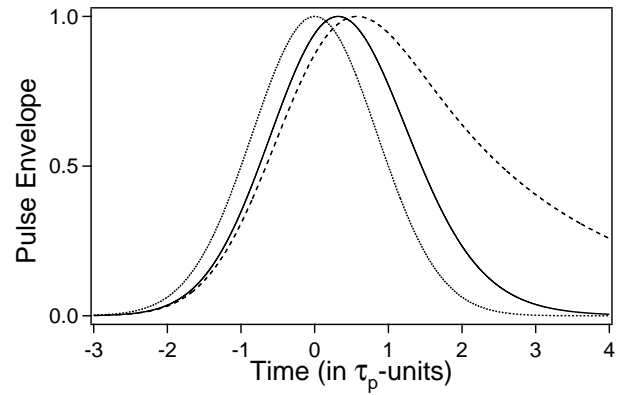} \caption{Same as Fig \ref{fig:RetardT2=00003D2T1} when $T_{2}\approx0$ (saturable
absorber). For the solid line $\tau_{p}/T_{1}=2.0$ , $\tau_{m}/\tau_{g}=0.79$
and $\tau_{m}/\tau_{p}=0.32$. For the dashed line $\tau_{p}/T_{1}=0.44$
, $\tau_{m}/\tau_{g}=0.31$ and $\tau_{m}/\tau_{p}=0.57$.\label{fig:RetardT2=00003D0}}

\end{figure}

For a saturable absorber ($T_{2}\approx0$) of same optical thickness
$\alpha\ell$, the maximal group delay, now attained for $I_{in}\approx4.8$,
is upgraded to $0.80T_{1}$ (see Fig.\ref{fig:NormalizedDelays})
whereas the transmission dynamics attains $15\,\mathrm{dB}$. Figure
\ref{fig:RetardT2=00003D0} shows the pulse envelopes obtained for
two different pulse durations. As expected, the maximum fractional
delay, obtained for $\tau_{p}\approx0.44T_{1}$ (dashed line), is
significantly larger than for the case of radiative relaxation (more
than two times larger). We however remark that the corresponding fall
of the pulse is considerably lengthened. An extensive study of the
pulse distortion in this particular system can be found in \cite{bm08}.

\section{Conclusion\label{sec:Conclusion}}

In their letter appeared in 1982 and soberly entitled {}``Linear
Pulse Propagation in an Absorbing Medium\textquotedblright{}, Chu
and Wong \cite{chu82} paved the way for the now-called fast-light
experiments. We have examined in the present article how the propagation
is modified when the absorbing medium is driven by a cw. Modeling
the absorbing medium as an ensemble of two-level atoms, we have more
specifically considered the case where both pulses and cw are on exact
resonance, a condition simply realized by pulse-modulating the cw
amplitude, with a low modulation index. This particular arrangement
eliminates the pulse-distortion associated with the first order variations
of the absorption and of the group delay versus frequency. The basic
result of our paper is the exact analytical expression of the transfer
function relating the Fourier transforms of the incident and transmitted
modulations for arbitrary values of the coherent and incoherent relaxation
times (Eq.\ref{eq:14}). It shows the importance of the effects resulting
from the depletion of the cw intensity along the propagation. When
the modulation is only harmonic (as in numerous experiments), it directly
gives the gain and the phase shift undergone by the modulation. They
significantly depart from those obtained with a unique probe field
(single sideband modulation). When the modulation is actually pulsed,
the transmission delay of the pulse center-of-mass, identified to
the group delay, is deduced from the transfer function by a simple
calculation of derivative. A remarkable point is that the group delay
is the sum of two terms, respectively proportional to the coherent
and incoherent relaxation times. These two terms being mainly of opposite
sign and depending differently on the cw intensity, this explains
why the transmission delay, strongly negative when the cw intensity
is low, may become (slightly) positive when the latter increases.
Finally the numerical determination of the pulse shapes confirms a
general property of the fast and slow light systems, namely that significant
advances or delays with moderate distortion can only be obtained in
media with a large transmission dynamics. Though our study is only
theoretical, it is illustrated for a realistic value of the optical
thickness, comparable to that actually used in the microwave experiment
reported in \cite{bs85}. In the optical domain, suitable optical
thickness and time-scale could probably be obtained by using an ensemble
of cold atoms. We expect that our theoretical work will stimulate
such an experiment.

\end{document}